\documentclass{article}

\usepackage{graphics}
 
\begin{document}
\newtheorem{thm1}{Theorem}
\newtheorem{thm2}[thm1]{Theorem}
\newtheorem{lemma}[thm1]{Lemma}
\newtheorem{lemma3}[thm1]{Lemma}
\newtheorem{lemma4}[thm1]{Lemma}
\newtheorem{prop3}[thm1]{Proposition}

\def\tr{ {\rm{Tr }}\,}
\def\pr{ {\rm{Pr }}}
\def\oti{{\otimes}}
\def\bra#1{{\langle #1 |  }}
\def\lb{ \left[ }
\def\rb{ \right]  }
\def\tilde{\widetilde}
\def\bar{\overline}
\def\*{\star} 

\def\({\left(}		\def\BL{\Bigr(}
\def\){\right)}		\def\BR{\Bigr)}
	\def\BBL{\lb}
	\def\BBR{\rb}
%
\newcommand{\qed}{\rule{7pt}{7pt}}
\def\E{{\mathbf{E} }}
\def\1{{ I }}
\def\bb{{\bar{b} }}
\def\ab{{\bar{a} }}
\def\zb{{\bar{z} }}
\def\zbar{{\bar{z} }}
\def\frac#1#2{{#1 \over #2}}
\def\inv#1{{1 \over #1}}
\def\half{{1 \over 2}}
\def\d{\partial}
\def\der#1{{\partial \over \partial #1}}
\def\dd#1#2{{\partial #1 \over \partial #2}}
\def\vev#1{\langle #1 \rangle}
\def\ket#1{ | #1 \rangle}
\def\proj#1{ | #1 \rangle \langle #1 |}
\def\rvac{\hbox{$\vert 0\rangle$}}
\def\lvac{\hbox{$\langle 0 \vert $}}
\def\2pi{\hbox{$2\pi i$}}
\def\e#1{{\rm e}^{^{\textstyle #1}}}
\def\grad#1{\,\nabla\!_{{#1}}\,}
\def\dsl{\raise.15ex\hbox{/}\kern-.57em\partial}
\def\Dsl{\,\raise.15ex\hbox{/}\mkern-.13.5mu D}
\def\b#1{\mathbf{#1}}
%
%
\def\th{\theta}		\def\Th{\Theta}
\def\ga{\gamma}		\def\Ga{\Gamma}
\def\be{\beta}
\def\al{\alpha}
\def\ep{\epsilon}
\def\vep{\varepsilon}
\def\la{\lambda}	\def\La{\Lambda}
\def\de{\delta}		\def\De{\Delta}
\def\om{\omega}		\def\Om{\Omega}
\def\sig{\sigma}	\def\Sig{\Sigma}
\def\vphi{\varphi}
%
%
\def\CA{{\cal A}}	\def\CB{{\cal B}}	\def\CC{{\cal C}}
\def\CD{{\cal D}}	\def\CE{{\cal E}}	\def\CF{{\cal F}}
\def\CG{{\cal G}}	\def\CH{{\cal H}}	\def\CI{{\cal J}}
\def\CJ{{\cal J}}	\def\CK{{\cal K}}	\def\CL{{\cal L}}

\def\CM{{\cal M}}	\def\CN{{\cal N}}	\def\CO{{\cal O}}
\def\CP{{\cal P}}	\def\CQ{{\cal Q}}	\def\CR{{\cal R}}
\def\CS{{\cal S}}	\def\CT{{\cal T}}	\def\CU{{\cal U}}
\def\CV{{\cal V}}	\def\CW{{\cal W}}	\def\CX{{\cal X}}
\def\CY{{\cal Y}}	\def\CZ{{\cal Z}}

\def\rvac{\hbox{$\vert 0\rangle$}}
\def\lvac{\hbox{$\langle 0 \vert $}}
\def\comm#1#2{ \BBL\ #1\ ,\ #2 \BBR }
\def\2pi{\hbox{$2\pi i$}}
\def\e#1{{\rm e}^{^{\textstyle #1}}}
\def\grad#1{\,\nabla\!_{{#1}}\,}
\def\dsl{\raise.15ex\hbox{/}\kern-.57em\partial}
\def\Dsl{\,\raise.15ex\hbox{/}\mkern-.13.5mu D}
\def\beq{\begin {equation}}
\def\eeq{\end {equation}}
\def\to{\rightarrow}

\title{The private classical  capacity and 
quantum capacity of a quantum channel}

\author{I. Devetak\footnote{
The author is with the IBM T.J. Watson Research Center, 
Yorktown Heights, NY 10598, USA. This work was supported
in part by the NSA under the US Army Research Office (ARO), 
grant numbers DAAG55-98-C-0041 and DAAD19-01-1-06.}
}

  \date{\today} 
  \maketitle

\begin{abstract}
A formula for the capacity of a quantum channel for transmitting private 
classical information is derived. This is shown to be equal
to the capacity of the channel for generating a secret key, 
and neither capacity is enhanced by forward public classical communication.
Motivated by the work of Schumacher and Westmoreland on quantum privacy and 
quantum coherence, parallels between private classical information and quantum 
information are exploited to obtain an expression for the capacity of a
quantum channel for generating pure bipartite entanglement. The latter implies 
a new proof of the quantum channel coding theorem and a simple proof
of the converse. The coherent information plays a role in all of the 
above mentioned capacities.

\end{abstract}

\section{Introduction}
The correspondence between secret classical information and quantum information,
after having been part of quantum information folklore for many years,
was first explicitly studied by Collins and Popescu \cite{cp}.
The simplest example of this relationship is the ability to convert 
a maximally entangled Bell state 
$\ket{\Phi^+} = {\frac{1}{\sqrt{2}}}(\ket{0} \ket{0} + \ket{1} \ket{1})$
shared by Alice and Bob into a secret classical key by local bilateral measurements 
in the $\{ \ket{0}, \ket{1} \}$ basis. Since the initial state is pure
and hence decoupled from the ``environment'', so is the 
information about the measurement outcomes.
The converse direction does not hold in the literal sense: 
there is no way to recover the entanglement once the measurement has been made.
However, given a quantum resource such as a quantum channel,
it is conceivable that a 
secret key generating protocol could be converted  into a (pure) entanglement 
generating protocol by performing all the steps ``coherently'' \cite{charlie},
e.g. replacing  probabilistic mixtures by quantum superpositions.
The connection has been exploited in one direction by Shor and Preskill 
\cite{shorpres} in proving the secrecy of the BB84 \cite{BB84} quantum key
distribution protocol by reduction from the entanglement-based protocol of Lo and 
Chau \cite{lochau} via Calderbank-Shor-Steane (CSS) \cite{CSS} codes. 
In a different context, an equivalence has recently been 
established  between the noise thresholds for certain two-way protocols for
secret key distillation and  entanglement distillation \cite{amg, amg2}.

The motivation for the present work is a paper by Schumacher and 
Westmoreland
\cite{sw1} in which an information theoretical approach to
secret key generation is taken. 
They invoke standard classical results on wire-tap channels 
\cite{wyner, ck2, ac1, maurer} to argue heuristically that the 
capacity of a noisy quantum channel $\CN$ for generating a secret key should be
lower bounded by the coherent information $I_c(\rho, \CN)$ 
\cite{sn, lloyd} of the channel with respect to an arbitrary input 
density operator $\rho$. Our first main result is an exact expression
for the channel capacity  for secret key generation $K(\CN)$. 
At the time of writing \cite{sw1} it was only conjectured that the
coherent information played a role in the quantum capacity
$Q(\CN)$. The quantum capacity theorem was originally stated by Lloyd
\cite{lloyd}, who also provided heuristic arguments for its validity.
Other relevant early works  include \cite{sn, bns, bkn}. 
It is only recently that a rigorous direct coding theorem has been 
reported by Shor \cite{q} attaining the coherent information based 
upper bound of \cite{bns, bkn}.
Our second main result is a new proof of the direct coding theorem
via an entanglement generation protocol based, in turn, 
on our secret key generation protocol. Shor's proof \cite{q}
is based on random subspace codes (see also \cite{lloyd}).
Our code turns out to be related to a generalization of CSS codes,
which is not surprising given its cryptographic origin.
In addition we provide a new, simplified proof of the converse
theorem of \cite{bkn}, avoiding difficulties with restricting
the encoding to partial isometries.

It is necessary to introduce some notation for dealing
with classical-quantum resources \cite{lock}. 
Classical-quantum resources can be  static or dynamic.
A static bipartite classical-quantum resource, denoted
by $\{ c \,q \}$, is described by an \emph{ensemble} 
$E = \{\rho_x, p(x)\}$. 
If the indices $x \in \CX$ and distribution $p$ are associated with some 
random variable $X$, and the density operators $\rho^\CQ_x = \rho_x$ with 
some quantum system $\CQ$, the ensemble $E$ may be equated with
the \emph{classical-quantum system} $X \CQ$.
One may similarly have multipartite systems such as $U X \CQ$ 
(of the $\{ c \,c \, q \}$   type) and
$X \CQ \CE$ (of the  $\{ c\, q\,  q \}$  type) 
with more than one classical or quantum component.

A dynamic bipartite classical-quantum resource, 
denoted by $\{ c \rightarrow q \}$ is given by
a classical-quantum channel $W: x \mapsto \rho_x$,
or, alternatively, by the \emph{quantum alphabet} $\{\rho_x\}$.
Analogous to the static case, the channel $W$ may be equated with the 
\emph{conditional} quantum system $\CQ|X$. Indeed, $\rho^\CQ_x$
is the state of the quantum system $\CQ$ conditioned on the classical 
index being $x$.
Dynamic resources are similarly extended to more than two parties. 

A useful representation of static classical-quantum systems, which
we refer to as the ``enlarged Hilbert space'' (EHS) representation,   
is obtained by embedding the classical random variables into quantum
systems.
For instance,  our ensemble  $E$ corresponds to the density 
operator 
\beq
\rho^{\CA \CQ} = \sum_x p(x) 
\ket{x}\bra{x}^{\CA} \otimes \rho_x^{\CQ},
\label{cq}
\eeq
where $\CA$ is a dummy quantum system and
$\{\ket{x}: x \in \CX \}$ is an orthonormal basis for the Hilbert space 
$\CH_\CA$ of $\CA$. A static classical-quantum system may, therefore, be viewed 
as a special case of a quantum one.
The EHS representation is convenient for
defining various information theoretical
quantities for classical-quantum systems. The von Neumann entropy of a quantum 
system $\CA$ with density operator 
$\rho^\CA$ is defined as $H(\CA) = - \tr \rho^\CA \log \rho^\CA$.
For a bipartite quantum system $\CA \CB$ define the conditional von 
Neumann entropy 
$$
H(\CB| \CA) = H(\CA \CB) - H(\CA),
$$
and quantum mutual information
$$
I(\CA; \CB) = H(\CA) + H(\CB) - H(\CA \CB) = H(\CB) - H(\CB| \CA),
$$
in formal analogy with the classical definitions.
For a tripartite quantum system $\CA \CB \CC$ define the quantum conditional
mutual information 
$$
I(\CA; \CB| \CC) = H(\CA| \CC) + H(\CB| \CC) - H(\CA \CB| \CC) 
                 = H(\CA\CC) + H(\CB\CC) - H(\CA\CB\CC) - H(\CC).
$$
A commonly used identity is the \emph{chain rule}
$$
I(\CA; \CB \CC) = I(\CA; \CB) + I(\CA; \CC | \CB).
$$
Notice that for classical-quantum correlations ($\ref{cq}$)
the von Neumann entropy $H(\CA)$ is just the
Shannon entropy $H(X) = - \sum_x p(x) \log \, p(x)$ of $X$.  
The conditional entropy $H(\CQ|X)$ is defined as $H(\CQ|\CA)$ and
equals $\sum_x p(x) H(\rho_x)$.
Similarly, the mutual information between $X$ and $\CQ$ 
is defined as  $I(X; \CQ) = I(\CA; \CQ)$. Notice that this
is precisely the familiar Holevo information \cite{holevo}
of the ensemble $E$:
$$\chi(E) = H\left( \sum_x p(x) \rho_x \right) - \sum_x p(x) H(\rho_x).
$$ 
Finally we need to introduce a classical-quantum analogue
of a \emph{Markov chain}.
A classical Markov chain $T \rightarrow X \rightarrow Y$
consists of correlated random variables $T$, $X$ and $Y$
whose probabilities obey 
$\pr\{Y = y | X = x, T = t\} = \pr\{Y = y | X = x\}$, which is to say
that $Y$ depends on $T$ only through $X$. Analogously we may define a
classical-quantum
Markov chain $T \rightarrow X \rightarrow \CQ$ associated with an ensemble
$\{  \rho_{tx}, p(t,x) \}$ for which $\rho_{tx} = \rho_x$. Such an object 
typically comes about by augmenting the system $X \CQ$ by the
random variable $T$ (classically) correlated with $X$ via a conditional
distribution $Q(t|x) = \pr\{T = t | X = x\}$. In the EHS representation
this corresponds to the state
\beq
\rho^{\CZ \CA \CQ} = \sum_x p(x) \sum_u Q(t|x) 
\ket{t}\bra{t}^{\CZ}  \otimes
\ket{x}\bra{x}^{\CA} \otimes \rho_x^{\CQ}.
\label{drei}
\eeq
We shall henceforth make liberal use of the concepts defined above
and their natural extensions.

The paper is organized as follows. In section 2 we
define and find expressions for the private information and key generation 
capacities $C_p(W)$ and $K(W)$, respectively, of a $\{c \rightarrow q q\}$ 
type channel $W$. We show that allowing a free forward public channel does
not help in either case. In section 3 these findings are applied to
a noisy quantum channel $\CN$ setting, yielding analogous capacities
$C_p(\CN)$ and $K(\CN)$. In section 4 we turn to the problem of entanglement
generation over the quantum channel $\CN$ and find the corresponding capacity
$E(\CN)$. This result is readily translated into an expression for the quantum
capacity $Q(\CN)$ in section 5. We conclude with open problems.

\section{Private information transmission and 
key generation over classical-quantum channels}

We begin by defining a general private information transmission
protocol for a $\{c \rightarrow qq\}$ channel from Alice to
Bob and Eve. The channel is defined by the map
$W: x \rightarrow  \rho^{\CQ \CE}_x$, with $x \in \CX$ and 
the $\rho^{\CQ \CE}_x$ defined on a bipartite quantum system 
$\CQ \CE$; Bob has access to $\CQ$ and Eve has access to $\CE$.
Alice's task is to convey, by some large number $n$ uses of the channel $W$
and unlimited use of a public channel (which both Bob and Eve have access to),
one of $2^{nR}$ equiprobable messages to Bob so that he can identify
the message with high probability while at the same time Eve
receives almost no information about the message.
The inputs to the composite channel $W^{\otimes n}$ are classical sequences 
of the form $x_1 \dots x_n \in \CX^n$, for which we use the shorthand
notation $x^n$ (not to be confused with the power operation).
The ouputs of $W^{\otimes n}$ are density operators living on some
Hilbert space $\CQ^n \CE^n$.
We formally define an $(n ,\epsilon)$ \emph{private channel code} of \emph{rate}
$R$ in the following way.
Alice generates a random variable $M$ which she can use for randomization, if
necessary. Given the classical message embodied in the random variable $K$
uniformly distributed on the set $[2^{n R}]:= \{1, 2, \dots,  2^{nR} \}$,
she sends the random variable $X^n = X^n(K, M)$ over the channel
$W^{\otimes n}$ and sends the random variable $S = S(K, M)$ through
the public channel. Bob performs a decoding POVM (based on the 
information contained in $S$)
on his system $\CQ^n$, yielding the random variable $Y$,
and computes his best estimate of Alice's message $L = L(Y, S)$.
We require 
\begin{eqnarray}
\Pr\{  K \neq L \}   &  \leq  &  \epsilon  \label{jen}, \\
I(K; S) & \leq & \epsilon \label{dva}, \\
I(K; \CE^n| S)  &  \leq  &  \epsilon. \label{tri} 
\end{eqnarray}
The second condition means that the public information $S$ is
almost uncorrelated with $K$ and the third implies via the Holevo bound
\cite{holevo} that, given the public information, there
is no measurement Eve could perform that would reveal more than
$\epsilon$ bits of information about $K$\footnote{As we shall
see, $\epsilon$ can be made to decrease {exponentially} in $n$.} .  
We  call the rate $R$ \emph{achievable} if for
every $\epsilon, \delta > 0$ and sufficiently large $n$
there exists an $(n, \epsilon)$ code of rate $R - \delta$.
The \emph{private channel capacity} $C_{p}(W)$ is the supremum of achievable rates 
$R$.

The above scenario should be contrasted with a secret key generation protocol, 
where Alice does not care about transmitting a particular message 
but only about establishing secret classical correlations with Bob, 
about which Eve has arbitrarily little information. The  
definition of an $(n, \epsilon)$ \emph{key generation code} 
is almost the same as that of a private channel code, with the
difference that now $K$ itself is a function of $M$.
The \emph{secret key capacity} $K(W)$ is similarly
given by the supremum of achievable rates $R$.

\begin{thm1} \label{t1}
\beq
  C_p(W) = K(W) =  \lim_{l \rightarrow \infty} \frac{1}{l}
\max_{T X^l} \{ I(T;\CQ^l) - I(T;\CE^l) \}, 
 \label{main}
\eeq
where $\CQ\CE | X$ is given by $W$ and
$T \rightarrow X^l \rightarrow \CQ^l \, \CE^l$ is a Markov chain.
\end{thm1} 

Note that the limit in equation (\ref{main}) indeed exists,
by standard arguments (see e.g. \cite{bns}, Appendix A).
It should be noted that the above formula does not
quite attain the ultimate goal of being effectively computable,
due to the ${l \rightarrow \infty}$ limit. This seems
to be a ubiquitous problem in quantum information theory,
and we shall encounter it two more times in this paper, 
namely, in theorem \ref{t2} and proposition \ref{p3}.

Proving that the right hand side of (\ref{main}) is
achievable is called the \emph{direct coding theorem}, whereas showing
that it is an upper bound is called the \emph{converse}.
It is obvious from our definition that $K(W) \geq C_p(W)$, since
any private channel can be used for generating a secret key.
Hence it suffices to prove the converse for $K(W)$ and achievability for 
$C_p(W)$.

\vspace {2mm}

\noindent  {\bf Proof of Theorem \ref{t1} (converse) }   \space \space
We shall prove that, for any $\delta, \epsilon > 0$ and 
sufficiently large $n$, if an $(n, \epsilon)$ secret key generation code has 
rate $R$  then 
$$
R - \delta \leq \frac{1}{n} 
\max_{T X^n} \{ I(T;\CQ^n) - I(T;\CE^n) \}.
$$
The proof parallels the classical one from \cite{ac1}.
Fano's inequality \cite{ct} says
$$
H(K|L) \leq 1 +  \Pr \{ K \neq L \} n R.
$$
Hence
\begin{eqnarray*}
n R & = &  H(K) \\
& = & I(K;L)  + H(K|L) \\
& \leq & I(K; L) + 1 + n \epsilon \log |\CX|.
\end{eqnarray*}
The last inequality follows from condition (\ref{jen}).
Furthermore,
\begin{eqnarray}
I(K; L)  &  \leq  & I(K; S \CQ^n)  \nonumber  \\
& = & I(K; S) + I(K; \CQ^n | S)  \nonumber  \\
&  \leq  & I(K; \CQ^n | S) - I(K; \CE ^n | S ) 
+ 2 \epsilon \nonumber   \\
&  =  & I(T; \CQ^n | S) - I(T; \CE ^n | S ) + 2 \epsilon,  
\end{eqnarray}
where  $T = K S$. The first inequality is a consequence
of the Holevo bound \cite{holevo} and the second inequality
follows from conditions (\ref{dva}) and (\ref{tri}). 
Since, without loss of generality,
$\epsilon \leq 
\frac {\delta}{6 \log |\CX|}$ and $n \geq \frac{2}{\delta}$,
\beq
\frac{1}{n}\left[ I(T; \CQ^n | S) - I(T; \CE ^n | S )\right]
\geq R - \delta,
\label{ema}
\eeq
with
 $S \rightarrow T \rightarrow X^n \rightarrow
\CQ^n \CE ^n$ a Markov chain. 
Since the left hand side can be written as the 
average of 
\beq
\frac{1}{n}\left[ I(T_s; \CQ^n_s) - 
I(T_s; \CE ^n_s )\right]
\label{sev}
\eeq
with respect to the distribution of $S$, and the Markov condition
$T_s \rightarrow X^n_s \rightarrow
\CQ^n_s \CE^n_s$ 
holds for each $s$,
choosing the particular value of $s$ that maximizes (\ref{sev})
proves the claim.\qed

\vspace{2mm}

For the direct coding theorem we shall need two lemmas.
The first is a quantum version of the Chernoff bound
from~\cite{aw}. 

\begin{lemma}[Ahlswede, Winter]
  \label{largedev}
  Let  $\xi_1,\ldots,\xi_\mu$ be independent identically distributed
  (i.i.d.) random variables with
  values in the algebra $B(\CH)$ of bounded linear operators
  on some Hilbert space $\CH$, which are bounded between
  $0$ and the identity operator $\1$. 
Assume that the expectation value $\E \xi_m =\theta \geq t\1$.
  Then for every $0<\eta<1/2$
  $$\Pr\!\left\{\frac{1}{\mu}\sum_{m=1}^\mu \xi_m \not\in[(1\pm\eta)\theta]\right\}
              \!\leq 2\dim{\cal H}\exp\!\left(\! -\mu \frac{\eta^2 t}{2\ln 2}\right)\!,$$
  where $[(1\pm\eta)\theta]=[(1-\eta)\theta;(1+\eta)\theta]$ is an interval
  in the operator order: $[A;B]=\{\xi\in{B}({\cal H}):A\leq \xi\leq B\}$.
  \qed
\end{lemma}

The second lemma is Winter's ``gentle operator'' lemma \cite{strong}. 
It says that a POVM element that succeeds on a state with high probability 
does not disturb it much.

\begin{lemma}[Winter]
  \label{tender}
  For a state $\rho$ and  operator $0\leq \Lambda\leq\1$,
  if $\tr(\rho \Lambda)\geq 1-\lambda$, then
  $$\left\|\rho-\sqrt{\Lambda}\rho\sqrt{\Lambda}\right\|_1\leq \sqrt{8\lambda}.$$
  The same holds if $\rho$ is only a subnormalized
  density operator.
  \qed
\end{lemma}

In the above, $\| A \|_1 = \tr \sqrt{A A^{\dagger}}$  
denotes the \emph{trace norm} of some operator $A$.
It is a norm in the sense that the \emph{trace distance} between
two operators $A$ and $B$, $\|A - B \|_1$,
satisfies the triangle inequality
\beq
\|A - C \|_1 \leq \|A - B \|_1  + \|B - C \|_1.
\label{ugao}
\eeq

\noindent  {\bf Proof of Theorem \ref{t1} (direct coding theorem) }   \space \space
We shall construct a private channel code that achieves
the expression (\ref{main})
without making use of the public channel from Alice to Bob.
Consequently, the public channel cannot increase $C_p(W)$ or
$K(W)$.
Fixing the random variable $X$ with  distribution $p$, 
our goal is first to show that a private information rate 
of $I(X;\CQ) - I(X;\CE)$ is achievable. 
We shall draw heavily on ideas from Winter's POVM compression paper
\cite{winter}. Define $\sigma_x = \tr_{\!\CQ} (\rho^{\CQ \CE}_x)$ and
$\omega_x = \tr_{\!\CE} (\rho^{\CQ \CE}_x)$, the local alphabets for 
Eve and Bob, respectively, and let $\omega = \sum_x p_x \omega_x$.

In what follows we shall assume familiarity with the notions of
typical sets $\CT^n_{X,\delta}$, typical subspaces $\Pi^n_{\CE, \delta}$ and
conditionally typical subspaces $\Pi^n_{\CE|X, \delta}(x^n)$. 
These are defined in  Appendix A for convenience.

Fixing $\delta > 0$, we have the following properties
(for $x^n  \in \CT^n_{X,\delta}$, where applicable) \cite{strong, qrst} :
\begin{eqnarray}
s := \Pr\{ X^n \in \CT^n_{X,\delta} \} & \geq &  1 - \epsilon \label{ulja}\\
{\tr \sigma_{x^n} \Pi^n_{\CE|X,\delta}(x^n)}  & \geq &  1 - \epsilon \label{bulja} \\ 
{\tr \sigma_{x^n} \Pi^n_{\CE, \, \delta(|\CX| + 1)}} & \geq &  1 - \epsilon 
\label{rulja}\\
{\tr \omega_{x^n} \Pi^n_{\CQ|X,\delta}(x^n)}  & \geq &  1 - \epsilon \label{bulja2} \\ 
{\tr \omega_{x^n} \Pi^n_{\CQ, \, \delta(|\CX| + 1)}} & \geq &  1 - \epsilon 
\label{rulja2}\\
\tr \Pi^n_{\CE, \, \delta(|\CX| + 1)} & \leq & \alpha^{-1}\label{hulja}  \\
 \Pi^n_{\CE|X,\delta}(x^n) \sigma_{x^n}  \Pi^n_{\CE|X,\delta}(x^n)
& \leq &  \beta \Pi^n_{\CE|X,\delta}(x^n)\label{ljulja} \\
\tr \Pi^n_{\CQ|X, \, \delta}(x^n) & \leq & \tilde{\beta}^{-1}\label{hulja2}  \\
\Pi^n_{\CQ, \, \delta(|\CX| + 1)}  \omega^{\otimes n} \Pi^n_{\CQ, \, \delta(|\CX| + 1)}
& \leq &  \tilde{\alpha} \Pi^n_{\CQ, \, \delta(|\CX| + 1)}.\label{ljulja2} 
\end{eqnarray}
Here $\alpha = 2^{-n[H(\CE) + c \delta]}$,
$\beta = 2^{-n[H(\CE|X) - c \delta]}$,
$\tilde{\alpha} = 2^{-n[H(\CQ) - c \delta]}$,
$\tilde{\beta} = 2^{-n[H(\CQ|X) + c \delta]}$
for some constant $c$
and $\epsilon = 2^{- n c' \delta^2}$ for some constant $c'$.
Define, for $x^n  \in \CT^n_{X,\delta}$,
\begin{eqnarray*}
\xi''_{x^n} & =  & \Pi^n_{\CE|X,\delta}(x^n)
\sigma_{x^n} \Pi^n_{\CE|X,\delta}(x^n), \\
\xi'_{x^n} & = & \Pi^n_{\CE,\delta(|\CX| + 1)} \xi''_{x^n} \Pi^n_{\CE,\delta(|\CX| + 1)}.
\end{eqnarray*}
Since $\sigma_{x^n}$ commutes with $\Pi^n_{\CE|X,\delta}(x^n)$,
$\xi''_{x^n} \leq \sigma_{x^n}$. From this, (\ref{bulja}) and (\ref{rulja}) 
\begin{eqnarray*}
\tr \xi'_{x^n}  & =  &  \tr \xi''_{x^n}  - \tr (\1 - \Pi^n_{\CE, \, \delta(|\CX| + 1)})
\xi''_{x^n} \\
& \geq & 1 - 2 \epsilon.
\end{eqnarray*}
Let $p'$ be the \emph{pruned} 
distribution $p^{\otimes n}$ with respect to the set $\CT^n_{X,\delta}$, namely
$$
p'(x^n) =  \left\{ \begin{array}{ll}
 \frac{p(x^n)}{s} & {x^n \in \CT^n_{X,\delta}} \\
0 & {\rm otherwise},
\end{array} \right.
$$
where $s$ is as defined in (\ref{ulja}).
Then $\tr \theta' \geq 1 - 2 \epsilon$, for
$$
\theta' = \sum_{x^n \in \CT^n_{X,\delta}} p'(x^n) \, \xi'_{x^n}.
$$
Let $\Pi$ be the projector onto the subspace spanned
by the eigenvectors of $\theta'$ with eigenvalue 
$\geq \epsilon \alpha$. 
By (\ref{hulja}), the support of $\theta'$ has dimension $\leq \alpha^{-1}$,
so eigenvalues smaller than $\epsilon \alpha$ contribute at most 
$\epsilon$ to $\tr \theta'$. Hence, $\tr \theta \geq 1 - 3 \epsilon$ 
for $\theta = \Pi \theta' \Pi$. Also let
$\xi_{x^n} =  \Pi \xi'_{x^n} \Pi$,
$\mu' =  2^{n[I(X;\CE) + 3 (c  + c' \delta) \delta] }$,
$\kappa' = 2^{n[I(X;\CQ) - I(X;\CE) - 5 (c + c' \delta) \delta] }$
and define $\mu' \kappa'$ i.i.d random variables $U_{km}$,
$m \in [\mu']$, 
$k \in [\kappa']$, each distributed according to $p'$.
Observe that $\theta = \E \xi_{U_{km}}$, where $\E$ denotes
taking expectation values with respect to the distribution $p'$.   
Define the event 
\beq
\iota_k = \left\{ \frac{1}{\mu'} \sum_{m = 1}^{\mu'} \xi_{U_{km}} \in 
[(1 \pm \epsilon)\theta ] \right\}. 
\label{hashe}
\eeq
According to lemma \ref{largedev},
$$
\Pr \{ {\rm not} \,\, \iota_k\} \leq 2 \tr \Pi 
\, \exp \left( - \mu' \frac {\epsilon^3 \alpha}{2 \beta \ln 2}   \right), \,\,\, 
\forall k.
$$
The right hand side, being a double exponential in $n$, can  be made
$\leq \epsilon{\kappa'}^{-1}$ for all $k$ and sufficiently large $n$. 
Now we shall argue that $\{ U_{km}$ is a good code for the 
$\{c \rightarrow q \}$ channel $\CQ|X$ \cite{strong} with high probability. 
It is a random code of size 
$2^{n[I(X;\CQ) - 2 (c + c' \delta) \delta]}$, and  each
codeword is chosen according to the pruned distribution $p'$.
The proof of the Holevo-Schumacher-Westmoreland theorem 
\cite{hsw}
involves choosing codewords according
to the distribution $p^{\otimes n}$ and can be easily modified
to 
work for $p'$ (see Appendix B). Consequently,  
the expectation of the average probability of error can be made to
decay exponentially with $n$:
$$\E \,p_e(\{ U_{km} \}) \leq 10 \epsilon.$$
Define the event 
\beq
\iota_0 = \left\{ p_e(\{ U_{km} \}) \leq \sqrt[4]{\epsilon^3} \right\}.
\label{stahr}
\eeq
Then Markov's lemma (from standard probability theory), 
according to which for any random variable $X$ and constant
$\gamma > 0$
$$
\Pr \{ X \geq \gamma \E X \} \leq \gamma^{-1}, 
$$
implies
$$
\Pr\{  {\rm not} \,\, \iota_0\}  \leq 10 \sqrt[4]{\epsilon}. 
$$
By construction,  
\beq
 \Pr\{ {\rm not} (\iota_0  \,\, \& \,\,  \iota_1 \,\, 
\dots \,\, \& \,\, \iota_{\kappa'}  ) \} \leq
 \sum_{k = 0}^{\kappa'} \Pr\{  {\rm not} \,\, \iota_k\} \leq \epsilon +
 10 \sqrt[4]{\epsilon},
\label{huk}
\eeq
so there  exists a \emph{particular} value 
$\{ u_{km} \}$ of $\{ U_{km} \}$,
for which $\iota_k$ holds for all $k = 0, \dots, \kappa'$
(in fact, we have shown this holds with probability 
$\geq 1 - \epsilon - 10 \sqrt[4]{\epsilon}$
for a \emph{randomly chosen}  $\{ u_{km} \}$).
For all $x^n \in \CT^n_{X,\delta}$ we have, by lemma \ref{tender},
\begin{eqnarray}
\| \sigma_{x^n} - \xi'_{x^n} \|_1 & \leq & 
\| \sigma_{x^n} - \xi''_{x^n} \|_1 + \|\xi''_{x^n}  - \xi'_{x^n}  \|_1 \nonumber \\
& \leq & \epsilon + \sqrt{ 16 \epsilon}. \label{udon}
\end{eqnarray}
Now, $\iota_k$ implies  
$$
\tr \left[ \frac{1}{\mu'} \sum_m \xi_{u_{km}} \right] \geq 1 - 4 \epsilon
$$
and hence, by (\ref{udon}) and lemma \ref{tender},
\begin{eqnarray}
\left\| \frac{1}{\mu'} \sum_{m = 1}^{\mu'}  \sigma_{\!u_{km}} 
- \theta \right\|_1 &  \leq &  
\frac{1}{\mu'} \sum_m \left\| \sigma_{u_{km}} - \xi'_{u_{km}} \right\|_1 +
\left\| \frac{1}{\mu'} \sum_m (\xi'_{u_{km}} - \xi_{u_{km}} ) \right\|_1 
+ \left\| \frac{1}{\mu'} \sum_{m = 1}^{\mu'}  \xi_{\!u_{km}} 
- \theta \right\|_1  \nonumber \\
& \leq & ( \epsilon + \sqrt{ 16 \epsilon}) + \sqrt{32 \epsilon} + \epsilon
= 2 \epsilon + 4 \sqrt{\epsilon} + 4 \sqrt{2\epsilon}.
\end{eqnarray}
So far we only have a bound on the \emph{average} error probability
for the channel code. We would like each individual codeword to
have low error probability. By (\ref{stahr}),
at most a fraction $\sqrt{\epsilon}$ of the codewords $u_{km}$
have  error probability $\geq \sqrt[4]{\epsilon}$.
Moreover, at most a fraction $\sqrt[4]{\epsilon}$
of the values of $k$ are such that a fraction 
$\geq \sqrt[4]{\epsilon} $ of the $u_{km}$ for that particular 
$k$ have  error probability $\geq \sqrt[4]{\epsilon}$.
We shall expurgate these values of $k$ from the code, without
loss of generality retaining the set $[\kappa]$ with 
$\kappa =  (1 - \sqrt[4]{\epsilon}) \, \kappa'$.
For each $k \in [\kappa]$,
order the $u_{km}$ according to increasing error probability,
and retain only the first $\mu = (1 - \sqrt[4]{\epsilon}) \, \mu'$ 
of them. This slight reduction in $\kappa'$ and $\mu'$ now ensures
that each codeword has error probability $\leq \sqrt[4]{\epsilon}$.
Since
\beq
\left\| \frac{1}{\mu'} \sum_{m = 1}^{\mu'}  \sigma_{\!u_{km}} 
-  \frac{1}{\mu} \sum_{m = 1}^{\mu}  \sigma_{\!u_{km}} \right\|_1 \leq 
2 \sqrt[4]{\epsilon},
\label{hoon}
\eeq
we now have 
\beq
\left\| \frac{1}{\mu} \sum_{m = 1}^{\mu}  \sigma_{\!u_{km}} 
- \theta \right\|_1 \leq
 2 \epsilon + 4 \sqrt{\epsilon} + 4 \sqrt{2\epsilon} +
2 \sqrt[4]{\epsilon} =: \epsilon'.
\label{anali}
\eeq
Note that this expurgation ensures that all the $u_{km}$ are distinct;
if they were not, the probability of error for a repeated codeword 
would be $\geq \frac{1}{2}$, a contradiction for sufficiently large $n$.
Defining
$$
\sigma_k = \frac{1}{\mu} \sum_{m = 1}^{\mu} \sigma_{\!u_{km}} 
$$
and 
\beq
\bar{\sigma} = \frac{1}{\kappa} \sum_{k = 1}^{\kappa} \sigma_k 
\label{ha}
\eeq
we have
\beq
\|\sigma_k - \bar{\sigma} \|_1  \leq 2 \epsilon', \,\,\,\, \forall k. 
\eeq
By Fannes' inequality (see e.g. \cite{nie&chuang}) we can estimate
\begin{eqnarray}
I(K; \CE^n) & = & 
\frac{1}{\kappa} \sum_{k = 1}^{\kappa} 
[H(\bar{\sigma}) - H(\sigma_k)]  \nonumber \\
& \leq & \eta(2 \epsilon') + 2 n  \epsilon'
\, \log \dim \CH_\CE,
\label{leff}
\end{eqnarray}
when $2 \epsilon' \leq \frac{1}{e}$ and where  
$\eta(x) = -x \log x$.
We are now in a position to describe our
private channel code. The random variable $M$ is uniformly distributed on 
$[\mu]$, the message $K$ is uniformly distributed on $[\kappa]$ and
the channel input is $X(K, M) = u_{KM}$. 
By construction, Bob can perform a measurement that 
correctly identifies the  pair $(k,m)$, and hence $k$, 
with probability $\geq 1 - \sqrt[4]{\epsilon}$.
The rate of the code is bounded as
$R \geq I(X;\CQ) - I(X;\CE) - 6 (c + c' \delta) \delta$, for sufficiently large $n$.
Equation (\ref{leff}) and $\epsilon = 2^{-c' n}$
ensures that $I(K; \CE^n)$ can be made arbitrarily small 
(indeed exponentially small in $n$) 
for sufficiently large $n$.
Notice that by simulating some channel $X|T$ in her lab, Alice
can effectively produce the $\CQ\CE|T$ channel for 
$T \rightarrow X \rightarrow \CQ\CE$; thus $I(T;\CQ) - I(T; \CE)$ is also 
achievable. The multi-letter formula (\ref{main}) follows from applying the above to 
the super-channel $W^{\otimes l}$. \qed

\vspace {2mm}

\noindent {\bf Remark} \,\,\, 
The classical analogue of theorem \ref{t1}, namely the
capacity $C_p(W)$ of a $\{c \rightarrow cc\}$ channel
$W = YZ|X$ was first discovered in \cite{wyner, ck2} for
a weaker notion of secrecy, and later strengthened in
\cite{maurer}. Our result implies a new proof of the classical direct coding 
theorem, using large deviation techniques instead of 
hashing/extractors as in the work of Maurer and collaborators 
\cite{maurer}.

\section{Private information transmission and 
key generation over quantum channels}

Now we shall apply these results to the setting where
Alice and Bob are connected via a  noisy \emph{quantum channel} 
$\CN: B(\CH_{\CP}) \rightarrow B(\CH_{\CQ})$.
Here $B(\CH_\CP)$ denotes the space of bounded linear operators
on $\CH_\CP$, the Hilbert space of the quantum system $\CP$. 
The channel $\CN$ is (non-uniquely) defined in terms of the 
operation elements $\{ A_i \}$, $\sum_i A_i^\dagger A_i = \1$, as
$$
\CN(\rho) = \sum_i A_i \rho A_i^\dagger.
$$ 
This representation is  exploited in Shor's proof of the quantum channel capacity
theorem \cite{q}. Here we take a different approach to noisy channels,
propagated by Schumacher and collaborators \cite{sn, bns}.
The channel is physically realized by
an isometry  $U_\CN: B(\CH_{\CP}) \rightarrow B(\CH_{\CQ \CE})$,
called an \emph{isometric extension} of $\CN$, 
which explicitly includes the unobserved \emph{environment} $\CE$.
\footnote{The standard formulation of \cite{sn} refers to a channel
$\CN: B(\CH_{\CQ}) \rightarrow B(\CH_{\CQ})$
with the same input and output Hilbert space. The channel is 
physically realized by appending an environment system $\CE$, wlog
initially in a pure state, applying a unitary 
operation $U^{\CE \CQ}$ on the joint system, and
tracing out $\CE$.
Here we adapt the slightly more general approach of \cite{bkn} in which 
the input and output Hilbert space of the channel may differ.}

We shall assume that the environment $\CE$ is completely under the control 
of the eavesdropper Eve, and the quantum system $\CQ$ is under Bob's
control.
Suppose Alice's initial density operator is given by 
$\rho^\CP$.
Defining $\omega^\CQ  = \CN(\rho^\CP) = \tr_{\!\CE} U_\CN (\rho^\CP) $
and $\sigma^\CE  = \tr_{\!\CQ}  U_\CN (\rho^\CP)$,  
the coherent information 
is defined as
$$
I_c(\rho^\CP, \CN) = H(\omega^\CQ) - H(\sigma^\CE).
$$
Note that, although there is an infinite family of $U_\CN$
corresponding to a given $\CN$, the coherent information
is independent of this choice \cite{sn}.
Since we are interested in transmitting private \emph{classical}
information, the most general protocol requires
Alice to prepend a $\{c \rightarrow q\}$  channel $\CP|X$ of her
choice (given by some alphabet $\{ \rho_x \} \in \CP^j$) 
to $j$ instances of $\CN$, for arbitrarily large $j$. 
This induces a $\{c \rightarrow q q\}$  channel
$\CQ^j \CE^j| X$, and we may now apply the results of the previous 
section. Combining the $l  \rightarrow \infty$ limit from equation (\ref{main}) 
with the $j \rightarrow \infty$ one and absorbing $T$ into $X$ gives 
$$
C_p(\CN)  =  K(\CN) = 
 \lim_{l \rightarrow \infty} \frac{1}{l}
\max_{X \CP} \{ I(X;\CQ^l) - I(X;\CE^l) \}.
$$
It is easily verified (see also section 4) that this may be rewritten as 
\beq
C_p(\CN)  =  K(\CN) =  \lim_{l \rightarrow \infty} \frac{1}{l}
\max_{\rho \in \CH_\CP^{\otimes l} }I_p (\rho, \CN^{\otimes l}),
\label{mane}
\eeq
where we introduce the \emph{private information} $I_p$: 
\beq
I_p(\rho, \CN) = I_c(\rho, \CN) -
 \min_{\{p(x), \rho_x \}} \left\{ \sum_x p(x) I_c(\rho_x, \CN)\, :  \,
\sum_x p(x) \rho_x = \rho \right\}.
\label{ibn}
\eeq
The above expression for $K(\CN)$ is almost implicit in \cite{sw1},
albeit without proof. Note that $I_p(\rho, \CN) \geq I_c(\rho, \CN)$ since any 
decomposition of $\rho$ into pure states
sets the expression minimized in (\ref{ibn}) to zero.
It is codes corresponding to $I_p(\rho, \CN) = I_c(\rho, \CN)$
that will be relevant for entanglement generation.
In Appendix C we give an example illustrating
the possibility of $I_p > I_c > 0$. 

\section{Entanglement generation over quantum channels}

In this section we apply the above results to the more
difficult problem of \emph{entanglement generation}
over quantum channels. The objective is for Alice and Bob
to share a nearly maximally entangled state on a 
$2^{nR} \times 2^{nR}$ dimensional Hilbert space, by
using a large number $n$ instances of the noisy 
quantum channel $\CN$. Before getting into details we
should  recall some facts about fidelities and purifications
(mostly taken from \cite{nie&chuang}). 
The fidelity of two density operators with respect to 
each other can be defined as\footnote{Our definition of fidelity
is the square of the quantity defined in \cite{nie&chuang}.}
$$F(\rho, \sigma) = \| \sqrt{\rho} \sqrt{\sigma} \|^2_1.$$
For two pure states $\ket{\chi}$, $\ket{\zeta}$ this
amounts to
$$
F(\ket{\chi}, \, \ket{\zeta}) = |\langle \chi \ket{\zeta}|^2.
$$


\begin{lemma} \label{resp}
Consider two  collections of orthonormal states 
$( \ket{\chi_j})_{j \in [N]}$ and $(\ket{\zeta_j})_{j \in [N]}$  such that  
$\langle \chi_j \ket{\zeta_j} \geq 1 - \epsilon$ for all $j$.
There exist phases $\gamma_j$ and $\delta_j$ such that
$$
\bra{\hat{\chi}} \hat{\zeta} \rangle \geq 1 - \epsilon,
$$
where 
\begin{eqnarray}
\ket{\hat{\chi}}& = & \frac{1}{\sqrt{N}} \sum_{j = 1}^{N} 
e^{i \gamma_j} \ket{\chi_j}, \\
\ket{\hat{\zeta}}& = & \frac{1}{\sqrt{N}} \sum_{j = 1}^{N} 
e^{i \delta_j} \ket{\zeta_j}.
\end{eqnarray}
\end{lemma}

\vspace{1mm}

\noindent  {\bf Proof}  \space \space
Define the Fourier transformed states
$$
\ket{\hat{\chi}_{s}} = \frac{1}{\sqrt{N}} \sum_{j = 1}^{N} 
e^{2 \pi i j s/N} \ket{\chi_j},
$$ 
and similarly define $\ket{\hat{\zeta}_{s}}$. 
It is easy to see that
$$
\frac{1}{N} \sum_{s = 1}^{N} \bra{\hat{\chi}_s} \hat{\zeta}_s \rangle =
\frac{1}{N} \sum_{j = 1}^{N} \bra{\chi_j} \zeta_j \rangle \geq 1 - \epsilon,
$$
hence at least one value of  $s$ obeys
$$
e^{i \theta_s} \bra{\hat{\chi}_s} \hat{\zeta}_s \rangle \geq 1 - \epsilon,
$$
for some phase $\theta_s$.
Setting $\gamma_j = 2 \pi j s/N$ and $\delta_j = \gamma_j + \theta_s$ satisfies the
statement of the lemma.

Moreover, a fraction $1 - \sqrt{\epsilon}$ of the 
values of $s$ satisfy
$$
e^{i \theta_s} \bra{\hat{\chi}_s} \hat{\zeta}_s \rangle \geq 1 - \sqrt{\epsilon},
$$
a fact we shall use in Appendix D.\,\,\,\qed

\vspace{1mm}

The following relation between 
fidelity and the trace  distance will be needed:
\beq
1 - \sqrt{F(\rho, \sigma)} \leq  \frac{1}{2} \| \rho - \sigma \|_1
\leq \sqrt{1 - F(\rho, \sigma)},
\label{omer}
\eeq
the second inequality becoming an equality for pure states.
A \emph{purification}  $\ket{\Phi_\rho}^{\CR \CQ}$ of 
a density operator $\rho^\CQ$ 
is some pure state living in an augmented quantum system $\CR \CQ$
such that $\tr_{\!\CR}( \ket{\Phi_\rho}\bra{\Phi_\rho}^{\CR \CQ}) = 
\rho^\CQ$. Any two purifications $\ket{\Phi_\rho}^{\CR \CQ}$
and $\ket{\Phi'_\rho}^{\CR \CQ}$ of $\rho^\CQ$ are related by 
some local unitary $U$ on the \emph{reference system} $\CR$
$$
\ket{\Phi'_\rho}^{\CR \CQ} = (U^{\CR}  \otimes \1^{\CQ}) 
\ket{\Phi_\rho}^{\CR \CQ}.
$$
A theorem by Uhlmann states that, for
a fixed purification $\Phi_\sigma$ of $\sigma$,
$$
F(\rho, \sigma)  = \max_{\Phi_\rho} F(\ket{\Phi_\rho}, \, \ket{\Phi_\sigma}).
$$
A corollary of this theorem is the \emph{monotonicity} property of
fidelity 
$$
F(\rho^{\CR \CQ}, \sigma^{\CR \CQ}) \leq 
F(\rho^{\CQ}, \sigma^{\CQ}),
$$ 
where $\rho^\CQ =  \tr_{\!\CR} \rho^{\CR \CQ}$
and $\sigma^\CQ =  \tr_{\!\CR} \sigma^{\CR \CQ}$.

Returning to the problem of entanglement generation, 
an $(n, \epsilon)$ code is defined as follows.
Alice prepares, without loss of generality, a \emph{pure} 
bipartite state $\ket{\Upsilon}^{\CA \CP^n}$ in her lab, 
defined on $\CH_\CA \otimes \CH_\CP^{\otimes n}$, 
$\dim \CH_\CA = \kappa$, and sends the $\CP^{n}$ 
portion of it through the channel.
Bob performs a general decoding quantum operation on the
channel output $\CD: B(\CH_\CQ^{\otimes n}) \rightarrow B(\CH_\CB)$,
$\dim \CH_\CB = \kappa$,
yielding the state 
\beq
\Omega^{\CA \CB} = 
[\1^\CA \otimes (\CD \circ \CN^{\otimes n})]
(\ket{\Upsilon}\bra{\Upsilon}^{\CA \CP^n}).
\eeq 
The rate of the code is  $R = \frac{1}{n} \log \kappa$. We require
$$
F(\ket{\Phi^\kappa}^{\CA \CB},\Omega^{\CA \CB}) \geq 1 - \epsilon,
$$
where 
$$
\ket{\Phi^\kappa}^{\CA \CB} =  
\sqrt{\frac{1}{\kappa}} \sum_{k= 1}^{\kappa} \ket{k}^\CA  \ket{k}^\CB
$$ 
is the standard maximally entangled state shared by Alice and Bob. 
We shall call a rate $R$ achievable  if for
every $\epsilon, \delta > 0$ and sufficiently large $n$ 
there exists an $(n, \epsilon)$ code of rate $R - \delta$.
The entanglement generating capacity $E(\CN)$
is given by the supremum of achievable $R$.

\begin{thm2} \label{t2}
Given the channel $\CN$,
\beq
E(\CN) =  \lim_{l \rightarrow \infty}
\frac{1}{l} \max_{\rho \in \CH_\CP^{\otimes l}} I_c(\rho, \CN^{\otimes l}).
 \label{main2}
\eeq
\end{thm2} 

\noindent {\bf Remark} \,\,\, Note
that $K(\CN) \geq E(\CN)$ is obvious since any
pure entanglement can be converted into a secret
key by performing a measurement in the $ \{ \ket{k} \}$ basis.
It is not clear from  formulas (\ref{mane}) and
(\ref{main2}) whether the inequality can be made  strict. 
We  return to
this issue in the final section. 

\vspace{2mm}

The converse theorem makes use of the following simple lemma \cite{bkn}.

\begin{lemma}
\label{bari}
For two states $\rho^{\CR \CQ}$ and $\sigma^{\CR \CQ}$
of a quantum system $\CR \CQ$ of dimension $d$
with fidelity $f = F(\rho^{\CR \CQ}, \sigma^{\CR \CQ})$,
$$
|\Delta H(\rho^{\CR \CQ}) - \Delta H(\sigma^{\CR \CQ})| \leq
\frac{2}{e} + 4 \log d \,  \sqrt{1 - f},
$$
where 
$$
\Delta H(\rho^{\CR \CQ}) = H(\rho^\CQ) - H(\rho^{\CR \CQ}). 
$$
\end{lemma}

\vspace{1mm}

\noindent  {\bf Proof}  \space \space
By the monotonicity of fidelity, $F(\rho^{\CB}, \sigma^{\CB}) \geq f$.
The lemma follows from a double application of Fannes' inequality 
\cite{nie&chuang} and (\ref{omer}).

\vspace{2mm}

\noindent  {\bf Proof of Theorem \ref{t2} (converse) }   \space \space
We shall prove that, for any $\delta, \epsilon > 0$ and 
sufficiently large $n$, if an $(n, \epsilon)$ code has rate
$R$  then $R - \delta \leq  
\frac{1}{n}  I_c(\rho, \CN^{\otimes n})$, 
where $\rho$ the restriction of $\ket{\Upsilon}$ to 
$\CH_\CP^{\otimes n}$.
Evidently, it suffices to prove this for $\epsilon \leq 
[\frac {\delta}{16 \log \dim \CH_\CP}]^2$ and $n \geq \frac{4}{e \delta}$.
The converse relies on the quantum data processing inequality,
which says that quantum post-processing cannot increase
the coherent information \cite{sn}. 
\begin{eqnarray*}
I_c(\rho, \CN^{\otimes n}) 
& \geq & I_c(\rho, \CD \circ \CN^{\otimes n}) \\
& = & \Delta H(\Omega) \\
& \geq & \Delta H(\ket{\Phi^\kappa} \bra{\Phi^\kappa}) -
 \frac{2}{e} - 8 n R \sqrt{\epsilon} \\
& \geq & n R - \frac{2}{e} - 8 n \log \dim \CH_\CP \sqrt{\epsilon}, 
\end{eqnarray*}
from which the claim follows. 
The first inequality is the data processing inequality and 
the second inequality is an application of lemma \ref{bari}.
\qed

\vspace{1mm}

\noindent  {\bf Proof of Theorem \ref{t2} (direct coding theorem) }  
\space \space
It suffices to demonstrate that a rate of $I_c(\rho, \CN)$ is achievable
for any $\rho \in \CH_\CP$. The regularized formula (\ref{main2})
is obtained by additional blocking. Following \cite{sw1}, 
consider the eigen-decomposition of $\rho$ into the orthonormal pure state ensemble
$\{ p(x), \ket{\phi_x} \}$, 
\beq
\sum_x p(x) \ket{\phi_x} \bra{\phi_x} = \rho.
\label{compost}
\eeq
The distribution $p$ defines a random variable $X$.
Let $U_\CN: B(\CH_{\CP}) \rightarrow B(\CH_{\CQ \CE})$
 be an isometric extension of $\CN$. Define the $\{ c \rightarrow qq \}$
channel $W: x \mapsto U_\CN  \ket{\phi_x}^\CP =: \ket{\phi'_x}^{\CQ\CE}$.
Define the local output density matrices  seen by Bob and Eve
by $\omega_x^\CQ = \tr_{\!\CE} (\ket{\phi'_x}\bra{\phi'_x}^{\CQ\CE})$
and $\sigma_x^\CE = \tr_{\!\CQ} (\ket{\phi'_x}\bra{\phi'_x}^{\CQ\CE})$, 
respectively, and the averages over $x$ by $\omega^\CQ$ and 
$\sigma^\CE$, respectively.
In section 2 we showed that for any $\delta$ 
there exists an $(n, \epsilon)$ code,  
defined by $\{ u_{km}: k \in [\kappa], m \in [\mu] \}$, of rate 
$\frac{1}{n} \log \kappa = I_c(\rho, \CN) - \delta$. Indeed \cite{sw1}
\begin{eqnarray*}
I(X; \CQ) - I(X; \CE) & = & H(\omega^\CQ) - \sum_x p(x) H(\omega_x^\CQ)   
- H(\sigma^\CE) + \sum_x p(x) H(\sigma_x^\CE) \\
& = &  H(\omega^\CQ) -  H(\sigma^\CE) \\
& = & I_c(\rho, \CN),
\end{eqnarray*}
since $H(\omega_x^\CQ) = H(\sigma_x^\CE)$ for all $x$.

In what follows we shall be dealing with blocks of length $n$
and use the abbreviated notation $\CQ$ for $\CQ^n$, etc.
Consider  sending the state $\ket{\phi_{km}}^\CP := \ket{\phi_{u_{km}}}^\CP$
through the isometric extension channel $U_\CN^{\otimes n}$
$$
\ket{\phi'_{km}}^{\CQ \CE} = U^{\otimes n}_\CN \ket{\phi_{km}}^\CP.
$$
Define 
$\sigma_{km}^{\CE} = \tr_{\!\CQ} (\ket{\phi'_{km}}
\bra{\phi'_{km}}^{\CQ \CE})$ and
$$
\sigma^\CE_k = \frac{1}{\mu} \sum_m \sigma^\CE_{km}.
$$
As shown in section 2, there exists a $\theta^\CE$ such that, for all $k$,
\beq
\| \sigma^\CE_k - \theta^\CE \|_1 \leq \epsilon.
\label{momir}
\eeq
In addition, there is a measurement Bob can perform on $\CQ$ 
that with probability $\geq 1 - \epsilon$ correctly identifies the index 
$km$. Since any measurement can be written as a unitary operation on a larger Hilbert
space (including some ancilla initially in a pure state) followed by
a von Neumann measurement on the ancilla, there exists a unitary $V^{\CQ \CB \CB'}$
such that 
$$
(\1^{\CE} \otimes V^{\CQ \CB \CB'}) 
\ket{\phi'_{km}}^{\CQ \CE} \ket{0}^\CB \ket{0}^{\CB'} = 
\ket{\psi_{km}}^{\CQ \CE \CB \CB'} 
$$
and
$$
F( \rho^{\CB \CB'}_{km}, \ket{k}^{\CB} \ket{m}^{\CB'} ) \geq 1 - \epsilon,
$$
for $\rho_{km}^{\CB \CB'} = \tr_{\!\CQ \CE} 
(\ket{\psi_{km}} \bra{\psi_{km}}^{\CQ \CE \CB \CB'})$.
By Uhlmann's theorem, for the purification 
$\ket{\psi_{km}}^{\CQ \CE \CB \CB'}$ of $\rho_{km}^{\CB \CB'}$ there
exists a ``purification'' $\ket{\chi_{km}}^{\CQ \CE} \ket{k}^\CB \ket{m}^{\CB'}$
of $\ket{k}^\CB \ket{m}^{\CB'}$
such that
\beq
\bra{\psi_{km}}^{\CQ \CE \CB \CB'} 
\ket{\chi_{km}}^{\CQ \CE} \ket{k}^\CB \ket{m}^{\CB'} \geq 1 - \epsilon.
\label{homer}
\eeq
Let $\tilde{\sigma}_{km}^{\CE} = 
\tr_{\!\CQ} (\ket{\chi_{km}} \bra{\chi_{km}}^{\CQ \CE})$.
Since $\ket{\psi_{km}}^{\CQ \CE \CB \CB'}$
is also a purification of $\sigma_{km}^{\CE}$,
we have, by   
 (\ref{omer}), the monotonicity of fidelity and  (\ref{homer}):
\begin{eqnarray*}
\|\sigma^\CE_{km} - \tilde{\sigma}^\CE_{km}\|_1 
& \leq & 2 \sqrt{1 - F(\sigma^\CE_{km}, \tilde{\sigma}^\CE_{km})} \\
& \leq & 2 \sqrt{1 - F( \ket{\psi_{km}}^{\CQ \CE \CB \CB'}, 
\ket{\chi_{km}}^{\CQ \CE} \ket{k}^\CB \ket{m}^{\CB'})} 
\leq 2 \sqrt{\epsilon}.
\end{eqnarray*}
Define 
\beq
\tilde{\sigma}^\CE_k = \frac{1}{\mu} \sum_m \tilde{\sigma}^\CE_{km}.
\label{umor}
\eeq
Then 
\begin{eqnarray}
\| \tilde{\sigma}^\CE_k - \theta^\CE \|_1 & \leq &
  \frac{1}{\mu} \sum_m \|\sigma^\CE_{km} - \tilde{\sigma}^\CE_{km}\|_1 +
 \| {\sigma^\CE}_k - \theta^\CE \|_1 \\
& \leq &  2 \sqrt{\epsilon} + \epsilon.
\end{eqnarray}
By  (\ref{omer})
\beq
F(\tilde{\sigma}^\CE_k, \theta^\CE) \geq 1 - 2 \sqrt{\epsilon} - \epsilon.
\label{djuka}
\eeq

Consider the  set of \emph{quantum codewords}
$\{\ket{\phi_k}\}$:
\beq
\ket{\phi_k}^\CP = \sqrt{\frac{1}{\mu}} \sum_{m} e^{i \gamma_{km}}
\ket{\phi_{km}}^\CP,
\label{ssss}
\eeq
with the phases $\gamma_{km}$ to be specified.
After transmission through $U_\CN^{\otimes n}$, adding the ancilla 
$ \ket{0}^\CB \ket{0}^{\CB'}$ and
applying  $V^{\CQ \CB \CB'}$,  $\ket{\phi_k}^\CP$ becomes
\beq
\ket{\psi_k}^{\CQ \CE \CB \CB'} = 
\sqrt{\frac{1}{\mu}} \sum_{m} e^{i \gamma_{km}}
\ket{\psi_{km}}^{\CQ \CE \CB \CB'}.
\label{hty}
\eeq
By (\ref{homer}) and lemma \ref{resp} we can choose the
phases $\gamma_{km}$ and $\delta_{km}$ so that
\beq
\bra{\psi_k}^{\CQ \CE \CB \CB'}
 \ket{k}^{\CB} \ket{\varphi_k}^{\CQ \CE \CB'} \geq 1 - \epsilon,
\label{smor}
\eeq
where
$$
\ket{\varphi_k}^{\CQ \CE \CB'}
=  \sqrt{\frac{1}{\mu}} \sum_{m} e^{i \delta_{km}} 
\ket{\chi_{km}}^{\CQ \CE} \ket{m}^{\CB'}.
$$
Note that $\tilde{\sigma}^\CE_k = 
\tr_{\!\CQ \CB'} (\ket{\varphi_k}\bra{\varphi_k}^{\CQ \CE \CB'})$,
as defined in (\ref{umor}).
Hence, fixing a purification $\ket{\Phi_\theta}^{\CQ \CE \CB'}$ of 
$\theta^\CE$,
for all $k \in [\kappa]$ there exists a unitary $U^{\CQ \CB'}_k$ such that 
(cf. \cite{sw2})
$$
 \bra{\Phi_\theta}^{\CQ \CE \CB'}[U_k^{ \CQ \CB'}  \otimes \1^\CE ] 
\ket{\varphi_k}^{\CQ \CE \CB'}
 \geq 1 -  2 \sqrt{\epsilon} - \epsilon,
$$ 
by applying Uhlmann's theorem to (\ref{djuka}).
Introducing  the ``controlled'' unitary 
$$
U^{\CB  \CQ\CB'} = \sum_k \ket{k} \bra{k}^\CB  \otimes
U_k^{ \CQ\CB'},
$$
the above may be rewritten as
$$
 \bra{\Phi_\theta}^{\CQ \CE \CB'} \bra{k}^\CB 
[U^{\CB \CQ \CB'}  \otimes \1^\CE ] 
\ket{k}^{\CB} \ket{\varphi_k}^{\CQ \CE \CB'}
 \geq 1 -  2 \sqrt{\epsilon} - \epsilon.
$$
Combining this with (\ref{smor}) gives
\beq
 \bra{\Phi_\theta}^{\CQ \CE \CB'} \bra{k}^\CB 
[U^{\CB \CQ \CB'}  \otimes \1^\CE ] 
 \ket{\psi_k}^{\CQ \CE \CB \CB'}
 \geq 1 -  4 \sqrt{\epsilon} - 4 \epsilon.
\label{hrana}
\eeq
We can now define our entanglement generating code. 
Alice prepares the state
\beq
\ket{\Upsilon}^{\CA \CP}
= \sqrt{\frac{1}{\kappa}} \sum_k \ket{k}^\CA \ket{\phi_k}^\CP,
\label{cvet}
\eeq
keeps the system $\CA$ and 
sends the system $\CP$  through the channel. 
Bob subsequently applies the decoding operator
\beq
\CD: \omega^\CQ \mapsto \tr_{\CQ \CB'}
\left[U^{\CB \CQ  \CB' } V^{\CQ \CB \CB' } 
(\omega^\CQ \otimes
 \ket{0} \bra{0}^\CB  \otimes  \ket{0}\bra{0}^{\CB'})
{V^{\CQ \CB \CB' }}^\dagger {U^{\CB \CQ \CB' }}^\dagger \right],
\label{dana}
\eeq
resulting in some state $\Omega^{\CA \CB}$, which
by (\ref{hrana}) and the monotonicity of fidelity obeys
\beq
F(\Omega^{\CA \CB}, \ket{\Phi^\kappa}^{\CA \CB}) 
\geq 1 -  4 \sqrt{\epsilon} - 4 \epsilon.
\label{fidelio}
\eeq
This concludes the proof of the direct coding theorem. \, \qed

\vspace{2mm}

\noindent {\bf Remark} \,\,\, Transforming a private channel
code into an entanglement generating one appears to work 
only for pure state decompositions  of $\rho$. 
Otherwise, the pure states $\ket{\phi'}$ become effectively
shared by {Alice}, Bob and Eve. The decoding operation 
$\CD$ would then involve performing joint operations on spatially
separated quantum systems belonging to Bob and Alice, which
cannot be accomplished in general without additional quantum resources.

\noindent {\bf Remark} \,\,\, Note the similarity between
(\ref{ssss}) and CSS codes \cite{CSS}. 
Indeed, here we have a coset-like decomposition of a
$\{c \rightarrow q \}$ ``error correction'' code of size $\kappa \mu$ 
into $\kappa$  $\{c \rightarrow q \}$ ``privacy amplification'' codes
of size $\mu$ 
(see \cite{nie&chuang} for a nice exposition of these concepts
in the context of the Shor-Preskill result \cite{shorpres}). 
The differences lie in that CSS codes have an 
additional algebraic structure and are composed of purely classical rather
than classical-quantum codes.  

\section{Quantum information transmission over quantum channels.}
Finally we arrive at our destination: recovering the formula for the 
quantum capacity $Q(\CN)$ of a quantum channel $\CN$. 
This quantity has been rigorously defined in \cite{bkn} and we
briefly review it here.
An $(n, \epsilon)$ code is defined
by an encoding operation $\CE: B(\CH) \rightarrow
B(\CH_\CP^{\otimes n})$ and a decoding 
operation $\CD: B(\CH_\CQ^{\otimes n}) 
\rightarrow B(\CH)$, such that 
\beq
\min_{\ket{\phi} \in \CH} F(\ket{\phi},
 (\CD \circ \CN^{\otimes n} \circ \CE) (\ket{\phi}\bra{\phi})) 
\geq 1 - \epsilon.
\label{uvjet}
\eeq
The rate of the code is given by $R = \frac{1}{n} \log \dim \CH$.
The quantum capacity of the channel ${Q}(\CN)$ is the supremum of all 
achievable $R$.

There is an alternative definition, in which the condition 
(\ref{uvjet}) and the definition of $R$ are replaced by
\beq
R = \max_{\rho \in \CH} \{ H(\rho) \, : \, F_e(\rho,
 \CD \circ \CN^{\otimes n} \circ \CE) \geq 1 - \epsilon  \}.
\label{buli}
\eeq
Here $F_e$ is the entanglement fidelity \cite{sn, nie&chuang}
$$
F_e(\rho, \CO) = F(\ket{\Psi}, (\1 \otimes \CO)(\ket{\Psi}\bra{\Psi})),  
$$
where $\ket{\Psi}$ is some purification of $\rho$ ($F_e$ is independent of
the particular choice of $\ket{\Psi}$).
We denote the corresponding capacity by $\tilde{Q}(\CN)$.
In \cite{bkn}, $Q(\CN)$ and $\tilde{Q}(\CN)$ were called the subspace
transmission and entanglement transmission capacities of the channel, 
respectively, and were shown to be equal.

It comes as no surprise that entanglement generation and 
entanglement transmission are closely related. This intuition 
is made rigorous by the following proposition.

\begin{prop3} \label{p3}
Given the channel $\CN$,
\beq
Q(\CN) = \tilde{Q}(\CN) =  E(\CN) = \lim_{l \rightarrow \infty}
\frac{1}{l} \max_{\rho \in \CH_\CP^{\otimes l}} I_c(\rho, \CN^{\otimes l}).
 \label{main3}
\eeq
\end{prop3} 

\vspace{1mm}

\noindent  {\bf Proof}  \space \space It is obvious that  
$E (\CN) \geq \tilde{Q}(\CN)$, 
since the quantum channel may be used to transmit
half of a maximally entangled state. Notice that this
fact in conjunction with the converse for theorem \ref{t2} 
yields a new and substantially simpler proof 
of the converse to the quantum channel capacity theorem (cf. \cite{bkn}).
To prove that $E (\CN) \leq \tilde{Q}(\CN)$ requires just a bit more work.
In \cite{bdsw, bkn} it was shown that 
$\tilde{Q}_{\rightarrow}(\CN) = \tilde{Q}(\CN)$, where
$\tilde{Q}_{\rightarrow}(\CN)$ is the quantum capacity
of a quantum channel $\CN$ enhanced by unlimited forward 
classical communication (cf. our corresponding result for private 
information transmission). Defining $E_{\rightarrow}(\CN)$ analogously, 
it now suffices to show  $E_{\rightarrow} (\CN) \leq \tilde{Q}_{\rightarrow}(\CN)$.
Indeed, any entanglement generated may be used in conjunction
with the forward  classical channel to perform quantum teleportation
 of the state $\rho$ \cite{tele}. More precisely (see section 4), Alice and Bob
generate the state $\Omega$ satisfying 
$$
f = F(\ket{\Phi^\kappa},\Omega) \geq 1 - \epsilon.
$$
They may perform a bilateral twirling operation \cite{bdsw} to transform
$\Omega$ into a Werner state
$$
\CT(\Omega) = \int dU (U \otimes U^*) \Omega  (U \otimes U^*) ^\dagger
= f \ket{\Phi^\kappa}\bra{\Phi^\kappa} + \frac{1 - f}{\kappa - 1}
(\1 - \ket{\Phi^\kappa}\bra{\Phi^\kappa} ), 
$$
which is now interpreted as being in the state 
$\ket{\Phi^\kappa}$ with 
probability $f$. Since teleporting a state $\rho$ living in a 
$\kappa$-dimensional Hilbert space $\CH$ via the maximally entangled 
$\ket{\Phi^\kappa}$ yields an entanglement fidelity of $1$, using 
$\CT(\Omega)$ instead will give an entanglement fidelity of at least 
$f \geq 1 - \epsilon$. Actually, one need not perform the full twirling
operation. The twirl is equivalent to applying some bilateral $U \otimes U^{*}$
chosen at random. Thus there exists a particular value of $U$ for which
the entanglement fidelity is $\geq f$. Furthermore, the
$U \otimes U^{*}$ is easily absorbed into $\ket{\Upsilon}$ and
$\CD$ of the entanglement generating protocol. 
Choosing $\rho$ to be maximally entropic proves the claim.
Thus $\tilde{Q}(\CN) =  E(\CN)$. \qed

\vspace{1mm}

\noindent {\bf Remark} \,\,\, It is possible to modify the proof of 
the direct coding part of theorem \ref{t2} to lower bound 
$\tilde{Q}(\CN)$ directly rather than via $E(\CN)$. 
This is done in Appendix D, where
we also show the existence of random entanglement transmission codes
of large blocklength $n$
with rate arbitrarily close to $I_c(\rho, \CN)$ and the nice property that 
the average density operator of the codewords is arbitrarily close to 
$\rho^{\otimes n}$.
   
\section{Discussion} We have defined and found expressions for
the private information transmission $C_p$ and secret key generation $K$
capacities for classical-quantum wire-tap channels $W$ and
quantum channels $\CN$. A subclass of the corresponding  protocols was made
``coherent'' to yield  entanglement generation and 
quantum information transmission protocols achieving the 
respective capacities $E(\CN)$ and $Q(\CN)$. Thus we have established 
a very important \emph{operational} connection  between quantum privacy and 
quantum coherence \cite{sw1}. 

Our results show that $C_p(W) = K(W)$, $C_p(\CN) = K(\CN)$ 
and $E(\CN) = Q(\CN)$. On the other hand,
it is obvious operationally, 
as well as from $I_p(\rho, \CN) \geq I_c(\rho, \CN)$, 
that $K(\CN) \geq E(\CN)$.
Although it is easy to find examples of strict
inequality between $I_c(\rho, \CN)$ and $I_p(\rho, \CN)$
for particular $(\rho, \CN)$ pairs (see Appendix C),
it is not clear whether this still holds when optimized over $\rho$ and
in the asymptotic sense of (\ref{mane}) and (\ref{main2}). 
In particular, we would like to know whether
there exist quantum channels which cannot be used for transmitting
quantum information, yet may be used to establish a secret key.
The ``static'' analogue of this question was answered recently in
in \cite{hhho} by providing an example
of a bipartite state with non-zero distillable secret key 
but zero (two-way!) distillable entanglement. 
It is natural to expect that a channel related to this state 
would demonstrate a separation between $K$ and $E$ \cite{john}.

Another open problem is whether the formula for $C_p(W)$ may be
single-letterized (as in the purely classical case \cite{ck2}) 
for general channels or at least certain classes of channels.
The same question is open for $C_p(\CN)$, whereas counterexamples
are known for $Q(\CN)$ \cite{dss}. A non-trivial class of channels 
exists (so-called \emph{degradable} channels)
for which $Q(\CN)$ is efficiently computable 
\cite{cqc}, and it can be shown that this property extends to
$C_p(W)$ (we do not know this to be true for $C_p(\CN)$).
More generally, we would like to be able to say something about
the convergence rate of the limits in equations (\ref{mane}) and 
(\ref{main2}).

A natural extension of the present work  would be to allow two-way
public/classical communication for key/entanglement generation. 
This enhanced auxiliary resource is known to improve the
capacities in both cases \cite{ac1, maurer, bdsw}, 
but it seems unlikely that reasonable
information-theoretical formulas exist in general. 

\vspace{1mm}
The results of section 2 were independently obtained by
Cai and Yeung \cite{cai}, see \cite{commit,wilmink}.

\vspace{1mm}

\noindent {\bf{Acknowledgments}} \, 
We thank C. H. Bennett and J. A. Smolin for useful discussions
and the former for pointing us to \cite{commit}.
Thanks also go to A. W. Harrow, D. W. Leung, K. El-Zein and J. Yard for comments on the
manuscript. Finally, we are indebted to A. S. Holevo and
D. Kretschmann for drawing our attention to a couple of technical errors in an 
earlier version of the paper.

\appendix

\section{Definitions of typical sequences and subspaces}

We shall list definitions and properties of
typical sequences and subspaces 
\cite{ck, nono, strong}. Consider
some general  classical-quantum system $U X \CQ$ in the state
defined by the ensemble $\{ p(u,x), \rho_{ux} \}$. 
$X$ is defined on the set $\CX$ 
and $U$ on the set $\CU$.
Denote by $p(x)$ and $P(x|u)$ the distribution of $X$ 
and conditional distribution of $X|U$, respectively.

For the probability distribution $p$ on the set $\CX$
define the set of \emph{typical sequences} (with $\delta>0$)
$$\CT^n_{p,\delta}=\left\{x^n:\forall x\ | N(x|x^n)- n p(x)|\leq
                              n {\delta} \right\},$$
where $N(x|x^n)$ counts the number of occurrences of
$x$ in the word $x^n=x_1\ldots x_n$ of length $n$.
When the distribution $p$ is associated with some random variable $X$
we may use the notation $\CT^n_{X,\delta}$. 

For the stochastic map $P: \CU \rightarrow \CX$ and
$u^n \in  \CU^n$ define the set of \emph{conditionally typical sequences} 
(with $\delta>0$) by
$$\CT^n_{P,\delta}(u^n) = \left\{x^n:\forall u,x \ | N((u,x)| (u^n,x^n)) 
- P(x|u) N(u|u^n)| \leq n{\delta} \right\}.$$
When the stochastic map $P$ is associated with some conditional 
random variable $X|U$  we may use the notation 
$\CT^n_{X|U,\delta}(u^n)$.

For a density operator $\rho$ on a $d$-dimensional Hilbert space $\CH$, 
with eigen-decomposition $\rho = \sum_{k = 1}^{d} \lambda_k \ket{k}\bra{k}$ 
define (for $\delta>0$) the \emph{typical projector} as
$$
\Pi^n_{\rho,\delta}=\sum_{k^n\in{\CT}^n_{R,\delta}}
\ket{k^n}\bra{k^n}.
$$
When the density operator $\rho$ is associated with some quantum  
system $\CQ$ we may use the notation 
$\Pi^n_{\CQ,\delta}$.

For a collection of states ${\rho}_u$, $u \in \CU$, and
$u^n\in \CU^n$ define the \emph{conditionally typical projector} as
$$\Pi^n_{\{\rho_u\},\delta}(u^n)=\bigotimes_u 
                              \Pi^{I_u}_{\rho_u,\delta},$$
where $I_u=\{i:u_i=u\}$ and $\Pi^{I_u}_{\rho_u,\delta}$
denotes the typical projector of the density operator ${\rho}_u$
in the positions given by the set $I_u$ in the tensor product of $n$ factors.
When the $\{\rho_u \}$ are associated with some conditional classical-quantum 
system $\CQ|U$ we may use the notation 
$\Pi^n_{\CQ|U,\delta}(u^n)$.

\section{The modified HSW theorem}
Define $\nu = 2^{n[I(X;\CQ) - 2(c + c' \delta) \delta]}$ i.i.d. random variables 
$\{ U_s\}$,
$s \in [\nu]$, according to the pruned distribution $p'$. 
We shall show that this random set can be made
into a HSW code for the $\{c \rightarrow q \}$ channel $\CQ|X$ with low
probability of error. More precisely, we shall construct a decoding POVM
$\{Y_s\}$ such that 
$$
\E p_e(\{U_s \}) = \E  \tr (\omega_{U_s} (\1 - Y_s)) \leq 10 \epsilon.
$$
We shall need the following lemma due to  Hayashi and Nagaoka \cite{hayashi}:  
For any operators $0 \leq S \leq \1$ and $T \geq 0$
\beq
\1 - (S + T)^{-\frac{1}{2}} S (S + T)^{-\frac{1}{2}}
\leq 2 ( \1 - S) + 4 T.
\label{hn}
\eeq
The $\{ Y_s \}$ are constructed as follows
$$
Y_s = (\sum_{t} \Lambda_{U_t})^{-\frac{1}{2}} \Lambda_{U_s}
 (\sum_{t} \Lambda_{U_{t}})^{-\frac{1}{2}} 
$$
with
$$
\Lambda_{x^n} = \Pi^n_{\CQ, \, \delta(|\CX| + 1)} \Pi^n_{\CQ|X,\delta}(x^n)
\Pi^n_{\CQ, \, \delta(|\CX| + 1)}.
$$
Then, by (\ref{hn}),
\beq
p_e(\{U_s \}) \leq 2 (1 - \tr \omega_{U_s} \Lambda_{U_s}) +
4 \sum_{t \neq s} \tr \omega_{U_s} \Lambda_{U_t}.
\label{glaz}
\eeq
It is not hard to verify (cf. lemma 6 of \cite{hayashi}) that
for $x^n \in \CT^n_{X, \delta}$
$$
\tr \omega_{x^n} \Lambda_{x^n} \geq 1 - 3 \epsilon
$$
follows from (\ref{bulja2}) and  (\ref{rulja2}).
By (\ref{hulja2})
$$
\tr \Lambda_{x^n} \leq \tilde{\beta}^{-1}.
$$
Also note that
$$
\E \omega_{U_s} = \sum_{x^n} p'(x^n) \omega_{x^n} \leq 
(1 - \epsilon)^{-1} \omega^{\otimes n}, 
$$
so that, by (\ref{ljulja2}),
$$
 \Pi^n_{\CQ, \, \delta(|\CX| + 1)} \E  \omega_{U_s} \Pi^n_{\CQ, \, \delta(|\CX| + 1)}
\leq (1 - \epsilon)^{-1} \tilde{\alpha}\, \Pi^n_{\CQ, \, \delta(|\CX| + 1)}.
$$
Putting everything together, taking the expectation of (\ref{glaz}) and noting 
that $U_s$ and $U_t$ are independent for $s \neq t$, we have
\begin{eqnarray*}
\E p_e(\{U_s \}) & \leq & 6 \epsilon + 
4 (\nu - 1) \tr (\E \omega_{U_s} \E \Lambda_{U_t})  \\
& \leq &  6 \epsilon + 4 \nu (1 - \epsilon)^{-1} \tilde{\alpha}\, \tilde{\beta}^{-1}
\leq 10 \epsilon, 
\end{eqnarray*}
the last inequality coming from our choice of $\nu$.

\section{An example of $I_p(\rho, \CN) > I_c(\rho, \CN)$}

Consider a four dimensional Hilbert space $\CH_4$, with  
orthonormal basis $ \{ \ket{1}, \ket{2} , \ket{3} , \ket{4} \}$. 
Let $\Pi_{12}$ be the projector onto the space spanned by
$\ket{1}$ and $\ket{2}$, 
and let $\Pi_{34}$ be the projector onto the space spanned by
$\ket{3}$ and $\ket{4}$. 
Our channel\footnote{This channel was suggested to
us by J. A. Smolin.} $\CN: B(\CH_4) 
\rightarrow B(\CH_4)$ is given  by 
$$
\CN(\rho) = \Pi_{12} \rho \Pi_{12} + \CD_{34}( \Pi_{34} \rho \Pi_{34}),
$$
where $\CD_{34}$ is the completely depolarizing channel on 
the two dimensional subspace spanned by $\ket{3}$ and $\ket{4}$.
Defining $\pi_{12} = \frac{1}{2}  \Pi_{12}$ and 
$\pi_{34} = \frac{1}{2} \Pi_{34}$, it is easily verified that
\begin{eqnarray}
I_c(\pi_{12}, \CN) & = & 1 \nonumber\\
I_c(\pi_{34}, \CN) & = & -1.
\end{eqnarray}
Define, for some small positive $\epsilon$,
\begin{eqnarray}
\pi'_{12} & = & (1 - \epsilon) \pi_{12} + 
\epsilon \pi_{34}\nonumber\\
\pi'_{34} & = & (1 - \epsilon) \pi_{34} + 
\epsilon \pi_{12}\nonumber.
\end{eqnarray}
By continuity, for sufficiently small $\epsilon$ 
\begin{eqnarray}
I_c(\pi'_{12}, \CN) & > & 0 \nonumber \\ 
I_c(\pi'_{34}, \CN) & < & 0.
\end{eqnarray}
Since 
$$
\pi'_{12} = \epsilon \pi'_{34} 
+ \frac{1 - \epsilon -\epsilon^2}{2} \proj{1}
+ \frac{1 - \epsilon -\epsilon^2}{2} \proj{2}
+ 
\frac{\epsilon^2}{2} \proj{3} + \frac{\epsilon^2}{2} \proj{4},
$$
is a valid decomposition of $\pi'_{12}$,
it is readily seen that 
$$
I_p(\pi'_{12}, \CN) > I_c(\pi'_{12}, \CN) > 0.
$$

\section{The average density operator of random quantum codes}
In this section we show how to convert a subclass of the entanglement 
generation codes 
 described in section 4  into entanglement 
transmission ones of the same rate $I_c(\rho, \CN) - \delta$.
Then we construct \emph{random} entanglement transmission codes
of the same rate such that 
the average density operator of the codewords becomes arbitrarily close to 
$\rho^{\otimes n}$ for large enough blocklength $n$.

Alice is given the system $\CA'$, entangled with some reference system $\CA$
she has no access to, in some general state with Schmidt decomposition 
$$
\ket{\Psi}^{\CA \CA'} = \sum_k \alpha_k \ket{k}^\CA \ket{k}^{\CA'}.
$$
Her goal is transfer the entanglement with $\CA$ from her system $\CA'$ to
Bob's $\CB$. Notice that the  states $\ket{\phi_{km}}^\CP$ 
(and hence $\ket{\phi_{k}}^\CP$)
are mutually orthogonal. Consequently, there is an isometric encoding
$\CE$ defined by $\ket{k}^{\CA'} \mapsto \ket{\phi_{k}}^\CP$
which maps $\ket{\Psi}^{\CA \CA'}$ to 
\beq
\ket{\Upsilon}^{\CA \CP}
=  \sum_k \alpha_k \ket{k}^\CA \ket{\phi_k}^\CP,
\label{cvetic}
\eeq
now bearing a strong resemblance to (\ref{cvet}).
By following through the remaining steps of the proof of theorem \ref{t2}, 
it is easily seen that after applying 
the decoding operation $\CD$ given by (\ref{dana})
one arrives at (c.f. (\ref{fidelio})):
\beq
F(\ket{\Psi}, (\1 \otimes (\CD \circ \CN^{\otimes n} \circ \CE))(\ket{\Psi}\bra{\Psi}))
\geq 1 - 4 \sqrt{\epsilon} - 4 \epsilon.
\label{fidq}
\eeq
Choosing $\ket{\Psi}$ to be maximally entangled implies, via (\ref{buli}), 
an achievable entropy rate of $\frac{1}{n} \log \kappa = I_c(\rho, \CN) - \delta$. 
 
\vspace{2mm}

The  set $\CS = \{\ket{\phi_k} \}$ is sometimes referred 
to as the \emph{quantum code}. A  natural quantity to define is 
the \emph{quantum code density operator}
$$
\rho(\CS) = \frac{1}{\kappa} \sum_{k = 1}^\kappa \ket{\phi_k} \bra{\phi_k},
$$
i.e. the input to the channel $\CN^{\otimes n}$ as seen by
someone ignorant of the encoded state.
Little can be said about $\rho(\CS)$ for any particular quantum code $\CS$ given by 
our construction. However if we consider \emph{random} codes, a probabilistic
mixture of deterministic codes given by an \emph{ensemble} 
$\{ p_\beta, \CS_\beta \}$, we can make the average code density operator 
$$
\bar{\rho} = \sum_\beta p_\beta \CS_\beta
$$
be arbitrarily close to $\rho^{\otimes n}$. We shall show
this via a double randomization of our original protocol.

1. First recall that for fixed $k$ and fixed set 
$\{ \ket{\phi_{km}}\}_{ m \in [\mu] }$, 
the $k$th quantum codeword (\ref{ssss})
$$
\ket{\phi_k} = \sqrt{\frac{1}{\mu}} \sum_{m} e^{i \gamma_{km}}
\ket{\phi_{km}}
\label{s4}
$$
was chosen from one of $\mu$ Fourier states.
If they were all 
``$\epsilon$-good'' quantum codewords, in the sense of (\ref{smor}),
then picking $\ket{\phi_k}$ at random according to the uniform
distribution on the set of Fourier states, 
for each $k$, would result in a random code with 
average code density operator
$$
\bar{\rho} = \frac{1}{\kappa \mu} \sum_{k, m} \ket{\phi_{km}} \bra{\phi_{km}}.
$$
According to the proof of lemma \ref{resp}, 
a fraction $1 - \sqrt{\epsilon}$ of the Fourier states 
are $\sqrt{\epsilon}$-good codewords. The random code
in which each  $\ket{\phi_k}$ is uniformly
distributed over these $\sqrt{\epsilon}$-good codewords
has an average code density operator $\bar{\rho}$ for which
\beq
\| \bar{\rho} - \frac{1}{\kappa \mu} \sum_{(k, m) \in [\kappa] \times [\mu]}
\ket{\phi_{km}} \bra{\phi_{km}} \|
\leq 2 \sqrt{\epsilon}.
\label{tru}
\eeq
At the same time, equation (\ref{fidq}) must be modified to 
account for the codes being $\sqrt{\epsilon}$-good instead of
$\epsilon$-good:
\beq
F(\ket{\Psi}, (\1 \otimes (\CD \circ \CN^{\otimes n} \circ \CE))
(\ket{\Psi}\bra{\Psi}))
\geq 1 - 6 \sqrt{\epsilon} - 2 \epsilon.
\label{trou}
\eeq
This concludes the first layer of randomization.

2. Backtracking to section 2, by equation (\ref{huk}), 
choosing  the
$\{u_{km}\}_{k \in [\kappa'], m \in [\mu']}$
\emph{at random} according to $p'$, followed by expurgation, results in failure
with probability $\epsilon + 10 \sqrt[4]{\epsilon}$,
thus modifying the fidelity estimate (\ref{trou})
to $1 - 10 \sqrt[4]{\epsilon} - 6 \sqrt{\epsilon} - 3 \epsilon.$
This is the second layer of randomization.
The average code density matrix before the expurgation is 
(denoting by ${\bf E}$ the expectation value over 
the $\{u_{km}\}$)
$$
{\bf E} \left[ \frac{1}{\kappa' \mu'} \sum_{(k, m) \in [\kappa'] \times [\mu']}
\ket{\phi_{km}} \bra{\phi_{km}} \right] =
\sum_{x^n} p'(x^n) \ket{\phi_{x^n}} \bra{\phi_{x^n}},
$$
for which
$$
\| \sum_{x^n} p'(x^n) \ket{\phi_{x^n}} \bra{\phi_{x^n}} 
- \rho^{\otimes n}\| \leq 2 \epsilon.
$$
The expurgation itself  has a small
effect on the average code density operator:
$$
\left\|  \frac{1}{\kappa \mu} \sum_{(k, m) \in [\kappa] \times [\mu]} 
\ket{\phi_{km}} \bra{\phi_{km}} 
-  \frac{1}{\kappa' \mu'} \sum_{(k, m) \in [\kappa'] \times [\mu']} 
\ket{\phi_{km}} \bra{\phi_{km}} 
\right\| \leq 4 \sqrt[4]{\epsilon},
$$
implying 
$$
\left\| {\bf E} \left[
 \frac{1}{\kappa \mu} \sum_{(k, m) \in [\kappa] \times [\mu]} 
\ket{\phi_{km}} \bra{\phi_{km}}  \right]
- {\bf E} \left[
 \frac{1}{\kappa' \mu'} \sum_{(k, m) \in [\kappa'] \times [\mu']} 
\ket{\phi_{km}} \bra{\phi_{km}} \right]
\right\| \leq 4 \sqrt[4]{\epsilon}  ,
$$
and hence
$$
\left\| {\bf E} \left[ 
\frac{1}{\kappa \mu} \sum_{(k, m) \in [\kappa] \times [\mu]} 
\ket{\phi_{km}} \bra{\phi_{km}} \right]
-  \rho^{\otimes n}
\right\| \leq 4 \sqrt[4]{\epsilon} + 2 \epsilon.
$$
By (\ref{tru}),
$$
\left \| \bar{\bar{\rho}} -  {\bf E} \left[
 \frac{1}{\kappa \mu} \sum_{(k, m) \in [\kappa] \times [\mu]}
\ket{\phi_{km}} \bra{\phi_{km}} \right] \right \|
\leq 2 \sqrt{\epsilon}.
\label{tru2}
$$
where $\bar{\bar{\rho}}:= {\bf E}\,\bar{\rho}$ is the average
code density matrix of our doubly randomized protocol.
Hence
$$
\| \bar{\bar{\rho}} - \rho^{\otimes n}\| 
\leq 4 \sqrt[4]{\epsilon} + 2 \sqrt{\epsilon} + 2 \epsilon,
$$
which concludes the argument.

\vspace {5mm}

{\bf Biography} I. Devetak received his Ph.D. 
in Electrical Engineering from Cornell University in 2002.
Since then he has been a post-doctoral researcher at the
IBM T.J. Watson Research Center, Yorktown Heights, NY.
He will be joining the Electrical Engineering faculty at the
University of Southern California in January 2005.

\end{document}